# Examine Clinicians' Modification of Hedging Language in Ambient AI Documentation: A Comparative Study of AI Drafts and Final Notes

**Yiliang Zhou, MS[1], Yawen Guo, MISM[1], Di Hu, MS[1], Sairam Sutari MS[2], Emilie Chow MD[3], Steven Tam MD[3], Danielle Perret MD[4], Deepti Pandita MD[3], Kai Zheng, PhD[1]**
**[1]Department of Informatics, University of California, Irvine, Irvine, CA, USA;**
**[2]Institute for Clinical and Translational Science, University of California, Irvine, Irvine, CA, USA;**
**[3]Department of Medicine, University of California, Irvine, Irvine, CA, USA;**
**[4]Department of Physical Medicine & Rehabilitation, University of California, Irvine, Irvine, CA, USA**

**Abstract**
*Ambient AI documentation systems generate clinical note drafts that clinicians frequently revise before signing off into electronic health records, yet how these edits alter hedging language remains unclear. We conducted paired analysis of clinician-edited portions of ambient AI drafts and final notes to examine (1) whether these edits change the prevalence of hedging language, (2) whether these edits exhibit a systematic shift toward greater certainty or uncertainty, and (3) whether these changes in hedging prevalence and directionality differ by ambient AI vendors and clinical specialties. Among 62,811 paired note sections, hedging terms were more often introduced into previously non-hedged text than removed from previously hedged text, and post-edit text contained more hedging mentions than pre-edit text. Directionality analyses showed a significant overall tendency toward greater uncertainty in hedging-related replacement edits. Vendor and specialty analyses revealed substantial heterogeneity in hedging prevalence, pre-to-post changes in hedging mentions, and directionality.*

## Introduction

Uncertainty is ubiquitous in medicine, spanning the diagnostic process, prognostication, and treatment decision making [1–4]. Clinicians routinely communicate varying degrees of likelihood, frequency, and magnitude when interpreting incomplete or evolving information [1,5]. In clinical documentation, these expressions are often recognized as hedging language, written or spoken terms that signal uncertainty, ambiguity, or underspecification in relation to clinical interpretation or guideline-based decision making [5,6]. Hedging terms are recognized through lexical and syntactic cues (e.g. "may," "often," "cannot rule out") that softens commitment to a single, definitive interpretation [5,7]. Such hedging terms help convey differential clinical diagnoses, guide next-step evaluation, and frame decisions about whether to initiate, continue, or withhold treatment [1,6–9].

Hedging is a core mechanism for documenting uncertainty responsibility and avoiding premature diagnostic closure [7,8,10]. However, uncertainty that is difficult to tolerate or interpret may contribute to clinician burden and has been linked to distress and burnout related outcomes [2,7,11]. Uncertainty expressions may also affect patients, as studies indicate that communicating diagnostic uncertainty can influence satisfaction and perceived competence, depending on how clearly uncertainty is framed and contextualized [2,7,8,12,13]. When uncertainty is conveyed without sufficient explanation, it may be misinterpreted as clinician indecision or poor-quality care, rather than an accurate reflection of evolving evidence [10,14,15]. Importantly, the effects of uncertainty are not uniformly negative: when communicated clearly and empathetically, acknowledging uncertainty may even strengthen the clinician–patient relationship by supporting continuity of care, patient-centered communication, and trust [11,16]. In the era of widespread electronic health records (EHRs) usage and increasing visibility of clinical notes [17], characterizing how uncertainty is documented and how it changes across documentation workflows, has growing relevance for both care delivery and patient experience.

Federal policy initiatives accelerated the digitization and availability of clinical documentation in the United States. The Health Information Technology for Economic and Clinical Health (HITECH) Act of 2009 promoted nationwide adoption of EHRs, and the 21st Century Cures Act further advanced health data access and interoperability, including expanding patients' ability to view clinicians' notes through patient portals [18–20]. Ambient Artificial Intelligence (AI) documentation platforms, building on these infrastructural advances, leverage automatic speech recognition (ASR) and generative AI (GenAI) to transform patient-clinician conversations into clinical notes [21–24]. Evidence suggests that ambient AI documentation tools may streamline clinician workflow and reduce documentation burden [22,25–30]. Reported benefits include improvements in clinician well-being (e.g., lower work exhaustion and interpersonal disengagement) as well as reduced time spent on documentation [21,30,31]. In operational evaluations, ambient AI use has also been associated with measurable productivity gains, including higher same-day encounter closure rates and less after-hours work time [23,30,32]. Because ambient AI platforms generate note drafts from conversations, in which lay language is more common than medical jargon and standard clinical hedging may be less consistently expressed, it is important to examine the content of these drafts and how clinicians revise them.

Emerging work has begun to assess ambient AI documentation quality. Lukac et al. found that clinicians rated clinically significant inaccuracies as occurring "occasionally" [25]. Guo et al. reported that clinicians routinely revise ambient AI drafts, with edits largely aimed at improving clinical accuracy and specialty-specific specificity [33]. Similarly, clinicians in prior qualitative work suggested that ambient AI drafts should default to conservative certainty, noting that they "tend to hedge more in medicine" and anticipating that including hedging terminology by default could reduce the need for subsequent corrections [31]. They also characterized hedging as a communication and risk-management practice, noting that phrases such as "low suspicion" may serve as a form of medico-legal protection [31]. Despite growing interest in documentation quality, the role of clinician editing in shaping hedging language in ambient AI-generated notes remains insufficiently understood. In this study, we used paired draft–final note comparisons to examine how uncertainty is managed during revision of AI-generated text. We focused on whether clinician edits increase or decrease the use of hedging, whether they shift statements toward greater tentativeness or greater definitiveness, and how these patterns differ across vendors and specialties.

## Materials and Methods

### *Dataset Description*

From late 2023, University of California Health has piloted two commercial ambient AI documentation platforms in outpatient settings. Vendors are referred to as Vendor A and Vendor B to preserve contractual and institutional confidentiality. For each encounter, we extracted paired ambient AI–generated draft and the corresponding clinician-authored final note. Clinicians can choose to use up to four note sections within a note, including History of Present Illness (HPI), Physical Exam, Assessment & Plan (A&P), and Results. Then they review and revise the generated text before finalizing the note. Detailed study characteristics and data collection procedures have been described in prior studies [27]. The use of these notes was reviewed and approved by the University of California, Irvine Institutional Review Board (UCI IRB #7123).

### *Hedging Lexicon Curation*

To identify hedging language in note text, we compiled two hedging lexicons from prior work—one derived from a prior study and the other from the CoNLL-2010 Shared Task resources [5,34]. After merging the lists and removing duplicate entries, three reviewers (Y.Z, Y.G, D.H) independently annotated each remaining term for directional effect on certainty, assigning labels of -1 for greater certainty, 1 for greater uncertainty, and 0 for neutral. Disagreements were resolved using majority vote. We calculated Fleiss' kappa to evaluate interrater agreement for the three reviewers' annotations of term directionality.

### *Sentence-Level Alignment and Edit Decomposition*

We operationalized clinician edits by aligning draft and finalized note text at the sentence level using pairwise sentence similarity. Sentence pairs with similarity score $\geq 75$ were eligible for one-to-one matching; unmatched draft sentences were labeled as drops, and unmatched finalized sentences as inserts [35]. Matched pairs with similarity $\geq 97$ or identical normalized text were labeled as equal, whereas matched pairs with similarity from 75 to <97 were labeled as replacements. For replacement pairs, we applied a Myers-style token-level diff to identify deleted draft spans and inserted finalized spans [36].

### *Clinical NLP Pipeline for Hedging Terms Assertion*

We developed a rule-based clinical NLP pipeline using SciSpaCy to identify hedging term mentions in note text and assign assertion labels. We utilized NLTK to compute sentence and word counts for each note section [37]. Hedging detection was implemented with an EntityRuler configured for case-insensitive phrase matching, with patterns constructed from our hedging lexicon and matched spans labeled as hedge entities. The pipeline was applied to section-level text from both ambient AI draft notes and clinician finalized notes. For each section, we generated structured outputs summarizing text length, the presence and frequency of hedging language, and the extracted hedge mentions. For each paired draft–final section, we also assigned an uncertainty directionality label based on reviewers' annotation, reflecting whether the clinician's edits made the text more certain, more uncertain, or directionally unchanged.

***Hedging Prevalence Analysis***

We examined clinician modifications to ambient AI–generated draft text at the level of edited note spans. Table 1 presents the three span types considered in this study, deletion, replacement, and insertion. The table shows how each appears in ambient AI draft text and clinician finalized text. For analyses of paired draft-final comparisons, we focused primarily on replaced text spans, since these provided the most symmetric comparison unit between pre-edit and post-edit text. Specifically, pre-edit text was defined as the draft text within replacement spans, and post-edit text was defined as the corresponding clinician edited replacement text. For example, if a draft span such as "the patient has viral infection" was revised to "the patient likely has viral infection," the original span was treated as pre-edit replaced text and the revised span as post-edit replaced text. Inserted text was summarized descriptively, but was not included in primary paired draft-final statistical comparisons.

**Table 1**. Edit span types and illustrative examples

| **Span type** | **AI draft** | **Clinician finalized note** |
|---|---|---|
| Deletion span | "This patient has pneumonia" | N/A |
| Replacement span | "This patient has pneumonia" | "This patient likely has pneumonia" |
| Insertion span | N/A | "This patient has pneumonia" |

For hedging prevalence analyses, we summarized the number and proportion of edited sections containing hedging before and after clinician modification, along with raw hedging mention counts. We also reported total hedging mentions across pre-edit and post-edit text. To evaluate paired changes in hedging prevalence within replaced text, we used McNemar's test on binary indicators of hedging presence before versus after editing. We used a one-sided Wilcoxon signed-rank test to assess whether post-edit replaced text contained more hedging mentions than pre-edit replaced text.

***Directionality Analysis of Clinician Edits***

To examine the directional tendency of clinician edits, we used certainty-direction scores derived from hedging term changes. At the replacement level, we calculated addition and deletion directionality scores; the replace net score was defined as the replacement addition score minus the replacement deletion score. We summarized score distributions descriptively and categorized net scores as shifts toward more certain, more uncertain, or no directional shift. We then used a one-sample Wilcoxon signed-rank test to assess whether net directionality scores significantly greater than 0.

***Stratified Analyses by Vendor and Specialties***

To assess heterogeneity across clinical and implementation contexts, we stratified analyses by ambient AI vendors and by a broad healthcare provider grouping of primary care (PCP) versus specialty care. For between-group comparisons, we used chi-square tests for hedging prevalence and Mann–Whitney U tests for raw hedging mention changes and directionality scores. In addition, we examined differences across individual specialties using chi-square tests for hedging prevalence and Kruskal–Wallis tests for raw hedging mention changes and directionality scores. All analyses were conducted in *R* version 4.1.3, and statistical significance was defined as P value $< 0.05$.

**Result**

***Study Cohort and Edit-Span Characteristics***

Of 98,311 paired ambient AI drafts and clinician finalized note sections, 62,811 remained after excluding sections with empty provider text and those with missing clinician specialty. Across these retained paired sections, clinician edits involved 216,199 dropped sentences, 52,542 replaced sentences, and 165,939 newly inserted sentences. Of the included paired sections, 47,403 were generated using Vendor A and 15,408 using Vendor B. In addition, 32,518 sections were from primary care encounters, and 30,293 were from encounters in 32 other specialties. Across the 62,811 retained paired sections, the mean word count was 3.8 (SD, 10.10) for pre-edit replaced text, 5.45 (SD, 13.42) for post-edit replaced text, and 40.19 (SD, 86.22) for inserted text. Restricting the analysis to the 9,891 replaced sections that contained at least one hedging mention in either pre-edit or post-edit text, the mean word count was 12.75 (SD, 20.01) for pre-edit replaced text and 19.80 (SD, 25.26) for post-edit replaced text.

***Hedging Lexicon Descriptive Statistics***

After compiling hedging terms from two prior sources and removing duplicates, the final hedging lexicon contained 536 unique terms. Inter-rater agreement among the three reviewers was assessed using Fleiss' kappa and was substantial ($\kappa = 0.72$, $P < 0.001$). Among the 536 hedging terms, 386 were labeled as shifting language toward greater uncertainty, 46 as shifting language toward greater certainty, and 104 as neutral with respect to certainty (as shown in Table 2). These annotations indicated that the lexicon was weighted more heavily toward terms associated with greater uncertainty than toward terms associated with greater certainty.

**Table 2:** Distribution of directionality labels in the hedging lexicon

| **Directionality Label** | **Count of Hedging terms** | **Example** |
|---|---|---|
| Toward greater certainty | 46 | "100% chance", "definitely", "always" |
| Neutral | 104 | "conclusion", "atypical" |
| Toward greater uncertainty | 386 | "odds on", "likely", "maybe" |

***Changes in Hedging Prevalence Within Replaced Text***

Across all clinician modified text, clinicians removed 6,990 hedging mentions from replaced draft text and added 10,381 hedging mentions within replacement spans. In addition, inserted text introduced 77,981 new hedging mentions. Within replaced text pairs, 5,088 of 62,811 sections contained at least one hedging mention in pre-edit text, compared with 6,756 sections in post-edit text. Among sections that contained at least one hedging mention, the mean number of mentions was 1.37 (SD, 0.82) across 5,088 pre-edit positive sections and 1.54 (SD, 1.04) across 6,756 post-edit positive sections.

**Table 3:** Distribution of paired hedging presence states in replaced text

| **Paired State** | **Total number of Sections** | **Proportion of Total Sections (%)** |
|---|---|---|
| Neither | 52,920 | 84.25 |
| Both | 1,953 | 3.11 |
| Pre-edit only | 3,135 | 4.99 |
| Post-edit only | 4,803 | 7.65 |

The distribution of paired section-level hedging presence states within replaced text indicated that clinicians were more likely to introduce hedging into previously non-hedged spans than to remove it entirely from previously hedged spans (as shown in Table 3). At the section level, 52,920 replaced sections contained no hedging before or after editing, 4,803 transitioned from no hedging to at least one hedging mention after editing, 3,135 transitioned from hedging present before editing to no hedging after editing, and 1,953 contained hedging both before and after editing. McNemar's test showed a significant asymmetry in these paired transitions ($P < 0.001$), indicating that clinicians were more likely to introduce hedging into previously non-hedged replaced text than to eliminate hedging from previously hedged replaced text. Consistent with this pattern, the one-sided Wilcoxon signed-rank test showed that post-edit replaced text contained significantly more hedging mentions than pre-edit replaced text ($P < 0.001$). To further characterize hedging-related edits within replaced text, we decomposed replacement spans into paired before–after hedging terms.We identified 1,117 unique hedging replacement pairs, and an example of hedging term replacement pairs is provided in Table 4. These pairs captured specific lexical substitutions through which clinicians revised hedging language within replaced spans.

**Table 4.** Top 10 hedging term replacement pairs observed in replaced text spans, sorted by decreasing frequency

| Pre-edit Term | Post-edit Term | Freq. |
|---|---|---|
| likely | possibly | 71 |
| or | if | 45 |
| consider | if | 28 |
| if | or | 26 |
| possible | or | 21 |
| likely | possible | 20 |
| may | or | 20 |
| or | possible | 20 |
| potential | or | 20 |
| if | consider | 19 |

***Directionality of Hedging-Related Replacement Edits***

To characterize the directional tendency of clinician hedging edits, we summarized net uncertainty scores derived from changes in hedging terms. Across the 62,811 retained paired sections, the mean replace net score was 0.02 (SD, 0.44). Positive scores indicate shifts toward greater uncertainty, whereas negative scores indicate shifts toward greater certainty. Most sections had no hedging-related replacement edit. When the analysis was restricted to the 6,568 sections with directional hedging edits in replaced text, the mean replace net score increased to 0.23 (SD, 1.33), indicating a stronger overall tendency toward greater uncertainty within sections where hedging-related directional edits occurred. Among these sections, 3,440 were classified as shifting toward greater uncertainty, 2,516 toward greater certainty, and 612 as showing no directional shift (Table 5). A one-sample Wilcoxon signed-rank test showed that the replace net score was greater than 0 ($P < 0.001$).

**Table 5.** Distribution of hedging edit directionality in replaced text

| Direction | Num. of Sections (%) |
|---|---|
| More certain | 2,516 (4.01%) |
| More uncertain | 3,440 (5.48%) |
| No directional shift | 612 (0.97%) |
| No hedging-related edit | 56,243 (89.54%) |

***Differences by Ambient AI Vendors and Specialties***

When stratified by ambient AI vendors, Vendor A sections contained 5,570 pre-edit hedging mentions and 8,018 post-edit hedging mentions, whereas Vendor B sections contained 1,420 pre-edit hedging mentions and 2,363 post-edit hedging mentions. Among replaced sections with post-edit hedging, the mean number of post-edit hedging mentions was similar across vendors, at 1.55 (SD, 1.05) for Vendor A and 1.50 (SD, 1.01) for Vendor B. Post-edit hedging prevalence within replaced text was slightly higher for Vendor A than for Vendor B (10.9% vs 10.2%) (chi-square test, P = 0.018). The net change in hedging mentions between pre-edit and post-edit replaced text also differed by vendor (Mann–Whitney U test, P = 0.012).When restricted to sections with directional hedging-related replacement edits, the mean replace net score was 0.20 (SD, 1.33) for Vendor A and 0.34 (SD, 1.31) for Vendor B, indicating that both vendors showed a trend toward greater uncertainty, with a stronger shift observed for Vendor B (Mann–Whitney U test, P < 0.001). The distribution of hedging edit directionality within replaced text also differed between ambient AI vendors, as shown in Table 6.

**Table 6.** Distribution of hedging edit directionality in replaced text between ambient AI vendors

| Direction | Num. of Sections (%) from Vendor A | Num. of Sections (%) from Vendor B |
|---|---|---|
| More certain | 2,041 (4.31%) | 475 (3.08%) |
| More uncertain | 2,648 (5.59%) | 792 (5.14%) |
| No directional shift | 501 (1.06%) | 111 (0.72%) |
| No hedging-related edit | 42,213 (89.05%) | 14,030 (91.06%) |

Across clinical groups, PCP sections contained 3,390 pre-edit hedging mentions and 5,083 post-edit hedging mentions, whereas specialty sections contained 3,600 pre-edit hedging mentions and 5,298 post-edit hedging mentions. Among sections with post-edit hedging, the mean number of hedging mentions was 1.50 (SD, 1.00) in PCP and 1.57 (SD, 1.08) in specialty care. Specialty sections were slightly more likely than PCP sections to contain at least one post-edit hedging mention (11.1% vs 10.4%; chi-square test, P = 0.003). However, the net change in hedging mentions from pre-edit to post-edit replaced text did not differ significantly between PCP and specialty sections (Mann–Whitney U test, P = 0.38).

Replace-level directionality scores were similar between PCP and specialty sections. Among sections with directional hedging-related replacement edits, the mean replace net score was 0.24 (SD, 1.31) for PCP and 0.21 (SD, 1.34) for specialty care. This difference was not statistically significant (Mann–Whitney U test, P = 0.41), indicating no evidence that the directional tendency of hedging-related replacement edits differed between PCP and specialty sections. The distribution of hedging edit directionality within replaced text also differed between primary care and specialty care, as shown in Table 7.

**Table 7.** Distribution of hedging edit directionality in replaced text between primary care and specialty care

| Direction | Num. of Sections (%) authored by PCP | Num. of Sections (%) authored by Specialty Care Provider |
|---|---|---|
| More certain | 1,225 (3.77%) | 1,291 (4.26%) |
| More uncertain | 1,717 (5.28%) | 1,723 (5.69%) |
| No directional shift | 315 (0.97%) | 297 (0.98%) |
| No hedging-related edit | 29,261 (89.98%) | 26,982 (89.07%) |

When extending the analysis across individual specialties, we restricted specialty-level summaries and tests to specialties with at least 20 eligible sections for the relevant analysis. Under this restriction, we observed significant heterogeneity in all three domains examined. Post-edit hedging prevalence differed across specialties (chi-square test, P < 0.001). Raw hedging mention change also varied significantly by specialty (Kruskal–Wallis test, P < 0.001), and replacement-level directionality scores differed across specialties among sections with hedging-related replacement edits (Kruskal–Wallis test, P < 0.001); the 15 specialties with the highest mean replace net scores are shown in Table 8. Together, these findings suggest that both the prevalence and directional patterning of clinician hedging edits vary across specialty contexts.

**Table 8.** Top 15 Specialties by Mean Replace Net Score, sorted by decreasing frequency

| Clinical Specialty | Replace Net Score, Mean | Post-edit Mentions, Mean |
| --- | --- | --- |
| Dermatology | 0.49 | 1.58 |
| Rheumatology | 0.38 | 1.54 |
| Neurology | 0.34 | 1.84 |
| Neurosurgery | 0.30 | 1.37 |
| General Surgery | 0.25 | 1.56 |
| Gastroenterology | 0.22 | 1.43 |
| Hand Surgery | 0.21 | 1.49 |
| Orthopaedic Surgery | 0.17 | 1.38 |
| Obstetrics and Gynecology | 0.15 | 1.42 |
| Colon and Rectal Surgery | 0.07 | 1.47 |
| Ophthalmology | 0.07 | 1.50 |
| Physical Medicine and Rehab | 0.05 | 1.65 |
| Epilepsy | 0 | 1.50 |
| Urology | -0.14 | 2.11 |
| Endocrinology | -0.26 | 1.74 |

**Discussion**

Uncertainty and ambiguity are common in clinical notes and often reflect the inherently provisional nature of clinical reasoning [1–5]. Hedging can function not only as a linguistic marker of uncertainty, but also as a communication and risk-management strategy that helps clinicians qualify assessments and avoid overstating certainty [11,16]. Ambient AI documentation provides an opportunity to examine these processes directly, and using the paired AI-generated draft and the clinician edited final text we examined how clinician edits changed the prevalence and directionality of hedging language within modified text spans, and whether these patterns differed by vendor and clinical context. Overall, we found that clinician edits more often introduced hedging into replaced text than removed it and that directional hedging edits tended to shift wording toward greater uncertainty.Previous studies have shown that finalized notes are often shorter than AI-generated drafts [27]. In our study, although edited note sections overall contained more mentions of hedging terms after revision, paired replacement-span analyses showed that, within the text clinicians actively rewrote, hedging was more often added than removed. Directionality analyses further showed a more uncertain directional tendency in hedging-related replacement edits. These findings suggest that, within the portions of ambient AI drafts that clinicians actively rewrote, editing more often moved documentation toward greater qualification rather than toward more definitive phrasing. These patterns are also consistent with prior interview findings, in which clinicians noted that they “tend to hedge more in medicine” and described hedging as part of routine communication and risk-management practice [31].

Differences by ambient AI vendors suggest that hedging-related edit patterns may partly reflect the characteristics of the draft text clinicians receive. Although vendor was not a significant factor associated with overall edit intensity in prior work [27], in our analyses, post-edit hedging prevalence, net change in hedging mentions, and replacement-level directionality all varied by vendor, with Vendor B showing a bigger shift toward greater uncertainty. However, the overall pattern was consistent across both systems: clinicians more often revised replaced text toward greater uncertainty than toward greater certainty. This indicates that hedging recalibration is likely a cross-vendor phenomenon rather than a problem unique to one platform. From a design perspective, these findings suggest that hedging language should be generated carefully and with closer grounding in clinical context. Tentative phrasing should reflect whether a diagnosis is confirmed or still under consideration and whether supporting evidence from labs, imaging, pathology, or other findings is present. Better alignment between uncertainty language and the underlying encounter may reduce unnecessary clinician correction and improve the reliability of ambient AI-generated notes.

Across clinical groups, specialty care note sections were slightly more likely than PCP sections to contain at least one post-edit hedging mention. However, neither the net change in hedging mentions nor the directionality of hedging-related replacement edits differed significantly between PCP and specialty sections. These findings suggest that primary care and specialty care may share a broadly similar tendency to revise ambient AI draft text toward greater uncertainty expression. When the analysis was extended to individual specialties, significant heterogeneity emerged in hedging prevalence, raw hedging mention change, and replacement-level directionality scores. These findings complement prior work on specialty-level differences in overall edit intensity by showing that specialties may differ not only in how extensively ambient AI drafts are edited, but also in how hedging and uncertainty are modified within edited text [27]. Such differences may reflect variation in diagnostic ambiguity, specialty-specific documentation norms, or different thresholds for expressing uncertainty in the medical record. In turn, these patterns suggest that ambient AI documentation systems may benefit from more specialty-specific models and editing support, particularly in how uncertain findings, provisional diagnoses, and evidence strength are represented in draft notes.

Our study has several limitations. First, we focused primary paired analyses on were limited to pre-edit and post-edit replacement text, and did not directly compare deleted text with newly inserted text. Deleted text may reflect many motivations beyond hedging revision alone, including redundancy removal, stylistic editing, workflow preferences, or medico-legal considerations, making it difficult to interpret deleted and inserted spans as symmetric counterparts. We therefore plan to examine deletion and addition content in greater detail, including which semantic types are preferentially added, removed, or preserved during editing. Second, we did not account for clinician or patient sociodemographic characteristics, and differences among note sections, which may shape clinical decision-making and, in turn, influence how uncertainty is expressed and revised in documentation [38]. Third, we did not fit multivariable models to account for potential confounding among covariates, including the possibility that ambient AI vendors were used disproportionately across specialties or clinical settings. As a result, some observed differences by vendor or clinical group may reflect underlying case-mix differences rather than independent effects of the vendor or specialty. Future work would integrate richer data to support drill-down analyses across note sections, clinician groups, and patient groups, enabling more contextualized understanding of how clinicians revise uncertainty language and informing AI tool improvement.

**Conclusion**

This study conducted a paired analysis of clinician-edited portions of ambient AI draft notes to assess how edits changed hedging prevalence, whether they systematically shifted documentation toward greater certainty or uncertainty, and whether these patterns differed by ambient AI vendor and clinical specialty. Overall, clinician edits within replaced portions of ambient AI drafts more often introduced hedging than removed it, and hedging-related replacement edits showed a significant overall tendency toward greater uncertainty rather than greater certainty. By subgroup, vendor-level differences were significant in post-edit hedging prevalence, net hedging change, and edit directionality, whereas PCP-versus-specialty differences were limited for directionality, although individual specialty analyses revealed significant heterogeneity in hedging prevalence, mention change, and replacement-level directionality. Future investigation is needed to support drill-down analyses incorporating richer patient context, case-mix adjustment, and identification of semantic types preferentially added or removed during editing.